\documentclass[prx,twocolumn,tightenlines,aps,epsf,showpacs,superscriptaddress,longbibliography]{revtex4-1}
\usepackage[pdftex]{graphicx}

\usepackage{dcolumn}
\usepackage{bm}
\usepackage{epsfig}
\usepackage{latexsym} 
\usepackage{amsmath}
\usepackage{amssymb}
\usepackage{color}
\usepackage{array}
\usepackage{bbm}
\usepackage{color}
\usepackage{array}
\usepackage{cancel}
\usepackage[dvipsnames]{xcolor}
\usepackage[normalem]{ulem}
\usepackage[mathscr]{euscript}
\usepackage{grffile}

\usepackage[colorlinks=true,allcolors=blue]{hyperref}%

\usepackage{titlesec}
\titleformat{\paragraph}[runin]
        {\bfseries}
        {}
        {0.0em}
        {}
        [ -- ~]
\titlespacing*{\paragraph}{0pt}{4pt}{0pt}

\usepackage[normalem]{ulem} 

\newcommand{\<}{\langle}
\newcommand{\e}{\varepsilon}

\renewcommand{\>}{\rangle}
\renewcommand{\(}{\left(}
\renewcommand{\)}{\right)}
\renewcommand{\[}{\left[}
\renewcommand{\]}{\right]}
\renewcommand{\vec}[1]{\boldsymbol{#1}} 

\renewcommand{\d}{\partial}

\begin{document}
\title{Qubit-efficient simulation of thermal states with quantum tensor networks}

\author{Yuxuan Zhang}
\email{yuxuanzhang@utexas.edu}
\affiliation{Department of Physics, The University of Texas at Austin, Austin, TX 78712, USA}

\author{Shahin Jahanbani}
\affiliation{Department of Physics, The University of Texas at Austin, Austin, TX 78712, USA}

\author{Daoheng Niu}
\affiliation{Department of Physics, The University of Texas at Austin, Austin, TX 78712, USA}

\author{Reza Haghshenas}
\affiliation{Division of Chemistry and Chemical Engineering, California Institute of Technology, Pasadena, California 91125, USA}

\author{Andrew C. Potter}
\affiliation{Department of Physics and Astronomy, and Stewart Blusson Quantum Matter Institute,
University of British Columbia, Vancouver, BC, Canada V6T 1Z1}
\begin{abstract}
We present a holographic quantum simulation algorithm to variationally prepare thermal states of $d$-dimensional interacting quantum many-body systems, using only enough hardware qubits to represent a ($d$-1)-dimensional cross-section. This technique implements the thermal state by approximately unraveling the quantum matrix-product density operator (qMPDO) into a stochastic mixture of quantum matrix product states (sto-qMPS).
The parameters of the quantum circuits generating the qMPS and of the probability distribution generating the stochastic mixture are determined through a variational optimization procedure.
We demonstrate a small-scale proof of principle demonstration of this technique on Quantinuum's trapped-ion quantum processor to simulate thermal properties of correlated spin-chains over a wide temperature range using only a single pair of hardware qubits. 
Then, through classical simulations, we explore the representational power of two versions of sto-qMPS ansatzes for larger and deeper circuits and establish empirical relationships between the circuit resources and the accuracy of the variational free-energy.  
\end{abstract}
\maketitle

\section{Introduction}
Simulating quantum materials and chemistry are leading contenders for near-term applications of quantum computers. 
A key practical constraint is that the effective size (e.g. number of coherently-operable qubits) of quantum processors that can be built with existing technology is severely limited. 
Conventional quantum simulation methods \emph{directly} encode each binary degree of freedom being simulated (e.g. spin-state or electronic orbital occupation) into a hardware qubit, and are strictly limited by the available number of qubits.
By contrast, holographic quantum simulation techniques~\cite{kim2017holographic, foss2021holographic}, \emph{indirectly} encode the model into a compressed tensor network state (TNS) form and utilize regular mid-circuit measurement and qubit resets (MCMR) to simulate much larger and more complex models than can simultaneously fit into quantum memory. 
These holographic methods are ``qubit-efficient" in the sense that their memory (qubit number) requirements are set by the amount of entanglement and correlation in the state being simulated rather than by the system size.
As an extreme example, ground-states of gapped-spin chains in the thermodynamic limit can be simulated as quantum matrix product states (qMPS, i.e. MPS whose tensors are implemented by quantum circuits) using only a handful of qubits~\cite{foss2021holographic}.

Rapid recent progress has led to a suite of holographic quantum simulation algorithms~\cite{foss2021entanglement,foss2021holographic,niu2021holographic,anand2022holographic} for approximately preparing the ground-states and performing non-equilibrium dynamics. 
Yet, many tasks of practical importance require computation of thermal properties of materials at non-zero temperature (e.g. calculating heat capacity, magnetic-ordering temperatures, etc.). 
Here, classical methods, such as classical tensor network calculations can become intractable in many contexts such as high-dimensions, or low-temperature~\cite{kuwahara2021improved}, creating an opportunity for quantum tensor network algorithms that are not limited by bond-dimension.
Preparing a thermal state is also a prerequisite for simulating quantum dynamics starting from a finite-temperature initial state, for example, to compute the temperature dependence of electrical or thermal conductivity in electronic devices or battery materials, or to model thermally-activated chemical reaction kinetics 
\cite{wang2008quantum}. 
We note that, for this purpose, it may be sufficient to prepare only a rough approximation of the thermal state of the desired temperature since the ensuing dynamics will produce rapid thermalization to the true thermal state with the temperature set by the initial energy density.
Thermal-like states such as the thermo-field double (TFD) states also play an important role in studies of quantum gravity through the Ads/CFT correspondence~\cite{cottrell2019build} (we note that the term ``holographic" is also used in that context, but with a different meaning than in this work).

Hybrid-classical/quantum algorithms based on variational methods offer a promising route for near-term implementations on noisy intermediate-scale quantum (NISQ) processors. 
A variety of methods for preparing thermal states to prepare a thermal state have been proposed~\cite{martyn2019product,wu2019variational,chowdhury2020variational,mcardle2019variational,zoufal2021variational,motta2020determining,zoufal2021variational,zhu2020generation,francis2021many,sagastizabal2021variational,cotler2019quantum}, including some demonstrations in hardware~\cite{motta2020determining,zoufal2021variational,zhu2020generation,francis2021many,sagastizabal2021variational,cotler2019quantum}. These include variational methods for preparing: mixed states with product spectra~\cite{martyn2019product}, TFDs~\cite{wu2019variational,chowdhury2020variational,zhu2020generation,francis2021many}, and imaginary time-evolution~\cite{mcardle2019variational,motta2020determining,zoufal2021variational}. Additionally, non-variational approaches have been proposed that are related to: simulating dynamics of a system-bath coupled to a bath~\cite{poulin2009sampling} (which relies on quantum phase estimation sub-routines that are out of reach for NISQ applications) or performing virtual cooling~\cite{cotler2019quantum} (which can require large sampling overhead).
These methods all involve both directly encoding the system into hardware qubits, and introducing an extensive number of additional ancilla to represent copies of the system (e.g. for purifications) or to simulate a thermal bath.
This qubit overhead is potentially limiting for NISQ-era implementations where qubit resources are a precious commodity.
Further, several of these approaches rely on measuring the thermal or entanglement entropy of a quantum state, a task that generically has exponential-in-system-size sampling complexity~\cite{cramer2010efficient}.


In this paper, we introduce a qubit-efficient method to holographically simulate thermal states of many-body systems, which circumvents many of these challenges. As a proof-of-principle and with an eye towards near-term hardware demonstration, we focus on $1d$ systems. However, the method readily generalizes to higher dimensions, where classical tensor methods can become challenging (e.g. by treating a $d$-dimensional system as a $1d$ stack of $(d-1)$-dimensional cross-sections, or employing higher-dimensional isometric tensor network representations).

The starting point of our approach is to represent a $1d$ thermal state as a matrix-product density operator (MPDO), which we then unravel into an ensemble of quantum circuited generated MPS that can be stochastically sampled to reproduce properties of the thermal MPDO. Crucially, the sto-qMPS format can be used without introducing an ancillary copy (e.g. TFD) of the system~\cite{verstraete2004matrix,wu2019variational} to purify the thermal state, thereby enabling thermal states to be approximately prepared with the same number of qubits as a pure state. We dub the resulting unravelling stochastic quantum-circuit generated MPS (sto-qMPS).
Adopting notation appropriate for a system of $N$-qubits or spins-1/2, each with basis states $|0\>,|1\>$, 
the basic idea behind this approach is that qMPS circuits turn a fixed reference initial state (e.g. $|00000\dots 0\>$) of physical qubits into an MPS. In sto-qMPS, the same type of qMPS circuit is applied to a random initial state of the physical qubits $|\vec{n}\>$ (with $\vec{n}\in \{0,1\}^N$ for a system with $N$ spins-1/2), where $x$ is drawn from a probability distribution $P(x)$ that can be efficiently sampled classically, and the results are averaged over the ensemble of qMPS described by the interplay of $P(x)$ and the unitary operations making up the qMPS circuit.
As we will discuss, this amounts to an oversimplification of the Hamiltonian on the scale of the many-body level spacing, which nonetheless has been argued not to dramatically affect physically-important properties~\cite{martyn2019product}.

The simplest instantiation of the sto-qMPS simply draws the initial physical qubit configurations from a product distribution, a qMPS adaptation of the so-called product-state ansatz (PSA) introduced in~\cite{martyn2019product}. Using PSA, we implement small-scale proof-of-principle demonstrations of this technique on Quantinuum's system model H1 trapped-ion quantum processor, achieving a rough approximation (relative errors $\sim 15-20\%$ accurate approximations) of thermal states of correlated spin chain models in the thermodynamic limit using only 2 hardware qubits with no error mitigation. We then explore the variational power of the sto-qMPS ansatz through numerical simulations on 2-6 qubit circuits of varying depths and show that the sto-qMPS provides a good approximation to physical properties (free-energy, correlation functions, etc.) of thermal states over a broad range of temperature scales. We find that the difficulty of approximating thermal states in this form is non-monotonic in temperature, becoming simplest in the low- and very-high temperature regimes, and achieving a maximum at temperatures of the order of the characteristic scale of terms in the Hamiltonian. 
We then show that introducing spatial correlations into $P$, which we refer to as a correlated spectrum ansatz (CSA), can surmount intrinsic limitations of the PSA, and comment on various possible extensions and refinements.

\section{Formalism}
To set the stage for the holographic implementation, we begin with a brief review of the main concepts of classical matrix product states and matrix product density operators (MPDOs).
%
\begin{figure*}[t]
\centering
\includegraphics[width=\textwidth]{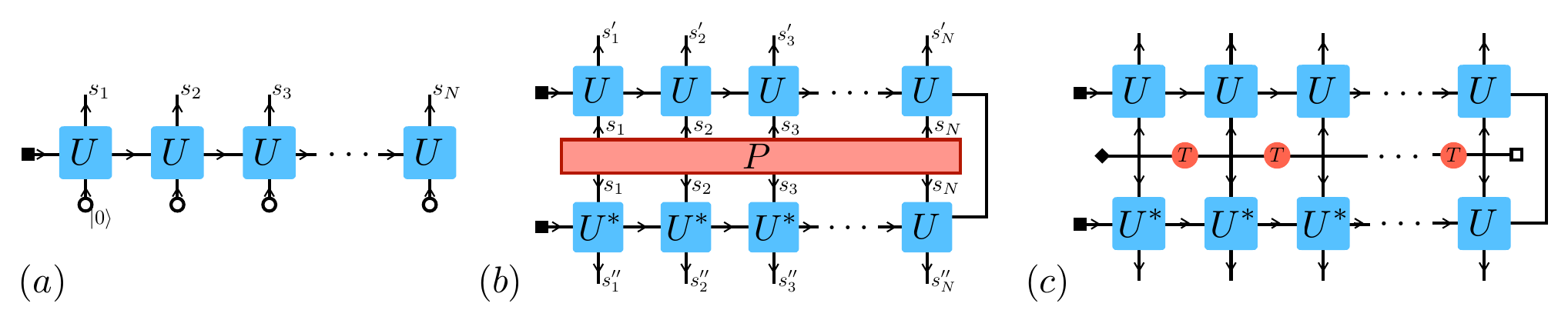}
\caption{
{\bf Graphical representations -- } of 
(a) quantum circuit generated matrix product (pure) state (qMPS), (b) stochastic qMPS (sto-qMPS) thermal state, and (c) a Boltzmann sto-qMPS where the stochastic mixture is generated from Boltzmann weights of a classical statistical mechanics model.
Blue rounded-corner squares are unitaries acting on physical (vertical lines) and bond (horizontal lines) qubits, with inputs and outputs indicated by arrows. Small squares indicate boundary vectors for bond qubits. (a) For qMPS, physical qubits are initialized to a fixed reference state (small open circles). (b) For sto-qMPS, the initial states of the physical qubits are sampled from a classical probability distribution, $P$, represented by a diagonal (in the computational basis) tensor. (c) In the Boltzmann sto-qMPS, $P$ is generated from Boltzmann weights of a classical statistical mechanics model with transfer matrix $T$, here, represented as a diagonal MPO with bond dimension $D$ equal to the physical dimension. Four-way line-junctions indicate that all four legs take the same index value. Filled and empty diamonds denote the left and right boundary vectors, $|\vec{1}\>$ (vector of all ones). 
}
\label{fig:MPDO}
\end{figure*}


\subsection{Matrix-Product- States and Density Operators} 
Any pure quantum state $|\Psi\>$ or mixed state $\rho$ can be expressed as a matrix-product state or density operator respectively. For example, this can be done by sequentially performing Schmidt decompositions between local sites, to write the wave-function amplitudes as a $1d$ tensor train:
\begin{align}
	|\Psi\> = \sum_{\{n_x\}_{x=1}^\infty} \ell^T A^{n_1}A^{n_2}\dots |n_1n_2\dots \>
\end{align} 
\begin{align}
	\rho = \sum_{n_1,m_1\dots } \ell^T B^{n_1,m_1}B^{n_2,m_2}\dots |n_1n_2\dots \>\<m_1m_2\dots|.
\end{align} 
Here, $n_x\in \{0,1,\dots D-1 \}$ label a basis of states for site $x$, and $A^{n_x}$, $B^{n_x,m_x}$ (for fixed $n_x,m_x$ label) are $\chi \times \chi$ matrices, and $\ell$ are $\chi$-dimensional vectors that determines the left boundary-conditions. 
Throughout this work, we will consider a semi-infinite $1d$ system with sites $x\in\{1,2,\dots \infty\}$.
The memory and cost of MPS (MPDO) computations scale with the bond-dimension, $\chi$, which is lower bounded by the bipartite entanglement (for $|\Psi\>$) or operator entanglement (for $\rho$) entropy across a cut through the bond.
For $1d$ short-range correlated, area-law entangled states~\cite{hastings2007area}, or thermal mixed states~\cite{hauschild2018finding}, one can truncate the entanglement spectrum to a system-size independent constant enabling efficient classical simulations of $1d$ gapped ground-states.
 Yet, many important cases remain out of reach for classical calculations. For example $2d$ and $3d$ systems can also be represented as MPS by treating them as a $1d$ stack of $(d-1)$-dimensional cross-sections. However, even for area-law entangled states, the required bond dimension grows exponentially with the \emph{cross-sectional area} -- which does not permit efficient classical simulation (although still offers substantial compression over exact simulation). In this context, quantum circuit generated tensor network methods may offer a significant advantage.

\subsection{Quantum-circuit-generated MPS (qMPS)}
Properties of any MPS in right-canonical form (RCF)~\cite{perez2006matrix} can be measured by sampling on a quantum computer and implementing its transfer-matrix as a quantum channel~\cite{gyongyosi2012properties} acting on $N_p=\log_2 D$ ``physical" qubits and $N_b=\log_2\chi$ bond qubits~(see Fig.~\ref{fig:MPDO}a for graphical representation). 
Each tensor $A$ is then embedded as a block of larger unitary operator $U_A$ acting on a reference initial state, $|0\>$, of the physical qubits: $A^n_{ij} = \<n|_p\otimes\<i|_b U_A |0\>_p\otimes |j\>_b$ where subscripts $p$ and $b$ respectively denote physical and bond qubits. The physical qubits can be measured in any desired basis (without measuring the bond qubits). 
The process is then repeated for each site in sequence from left to right. In this way, one can measure any product operator of the form $\prod_{x=1}^L \mathcal{O}_x$, which forms a complete basis for general observables. 
Crucially, once measured, the physical qubit for site $x$ can be reset to $|0\>$ and reused as the physical qubits for site $x+1$, enabling a small quantum processor to achieve quantum simulation tasks with sizes far larger than the number of qubits available ~\cite{foss2021holographic}.

To summarize, the qMPS procedure for sampling an observable of the form $\<\psi|\prod_{x=1}^L O_x|\psi\>$ is:
\begin{enumerate}
\item[0.] Prepare the bond qubits in a state corresponding to the left boundary vector $\ell$.
\item Reset the physical qubit for site $[x]$ in a fixed reference state, $|0\>$.
\item Perform a quantum circuit representing $U_A$ at site $[x]$, entangling the physical and bond qubits. \item Measure the physical qubit in the eigenbasis of $O_x$ and weight the measurement outcome by the corresponding eigenvalue of that observable. The bond-qubit register now corresponds to bond connecting sites $x$ and $x+1$.
\item Repeat steps 1-4 for $x=1\dots L$, and discard the bond-qubits~\footnote{In the last step, there is no reason to continue the chain beyond site $x=L$ where $L$ is the rightmost site where the observable has support: since we are preparing a state in RCF, the tensor contractions without operator insertions for $x>L$ simply multiply the $x\leq L$ network from the right by the identity vector, which corresponds to tracing out or discarding the bond qubits.}.
\end{enumerate}

Moreover, the entanglement spectrum of the bond-qubits in between sites $x$ and $x+1$ coincides with the bipartite entanglement spectrum of the physical MPS at that entanglement cut, further enabling measurement of non-local entanglement observables, as recently demonstrated experimentally~\cite{foss2021holographic}. The left boundary-vector $\ell$ is prepared by a unitary circuit from a fixed reference state. The right boundary conditions are not specified in this framework, which formally describes a semi-infinite wire (or equivalently simulating a length $L$ chain, but tracing observables over right boundary vectors~\cite{foss2021holographic}).

qMPS methods enable qubit-efficient access to a subset of MPS with exponentially-large bond dimension (in qubit number), including classically intractable cases such as $2d$ and $3d$ ground-states with symmetry-breaking~\cite{niu2021holographic} or (non-chiral) topological order~\cite{soejima2020isometric}, and finite-time quantum dynamics from any qMPS~\cite{foss2021holographic,chertkov2021holographic}.


\subsection{Quantum circuit generated MPDOs}
%
It is natural to ask whether holographic techniques can also be used to prepare thermal states as quantum-circuit generated MPDOs (qMPDO), for example, to simulate an infinite $1d$ thermal state using finite quantum memory.
One possibility would be to simulate a purification of the thermal state as a qMPS on an enlarged Hilbert space using extra ancilla qubits, for example to prepare a thermo-field double (TFD). 
Tracing out the ancillary degrees of freedom would give a mixed state for the physical qubits, which could potentially approximate a thermal state.
There are two main difficulties with this procedure. 
First, creating the purified thermal-state requires (at least) twice as many physical qubits as for a pure state.
Second, and more fundamentally, one must find an approximate representation of the purified qMPS tensors in terms of unitary circuits generated by local one and two-qubit gates.
 For ground-state preparation, this can be approximately done by choosing a variational circuit ansatz with circuit parameters $\vec{\theta}$ and minimizing $E(\vec{\theta}) = \<\psi(\vec{\theta})|H|\psi(\vec{\theta})\>$ with respect to the variational parameters, where $|\psi(\vec{\theta})\>$ is the qMPS generated by circuit parameters $\vec{\theta}$.
 There is a similar variational principle for thermal states: namely given a parameterized mixed-state, $\rho(\vec{\theta})$, then, the variational free energy $F(\vec{\theta}) = E(\vec{\theta})-TS(\vec{\theta})$ provides a variational upper-bound on the true free-energy, $F(\vec{\theta})\geq F_\text{exact}$. Here we have defined the variational-energy: $E(\vec{\theta})=\text{tr}(\rho(\vec{\theta}) H)$ and entropy: $S(\vec{\theta}) = -\text{tr}(\rho\log\rho)$.
While $E$ can be efficiently measured on a quantum computer for any mixed state, measuring the entropy $S$ generally has sampling complexity that grows exponentially in system size (using, for example, full-tomography~\cite{cramer2010efficient}, SWAP-trick~\cite{kliesch2014locality}, or randomized measurements~\cite{brydges2019probing}).

\subsection{sto-qMPS}
Here, we introduce an approach that evades both of these difficulties. We replace the task of sampling from a purified-mixed state, with sampling from a classical distribution of pure qMPS's, which we call a stochastic qMPS (sto-qMPS). 
Specifically, sampling observables in sto-qMPS proceeds similarly to that for a qMPS, except that instead of resetting the physical qubit(s) to a fixed reference state $|0\>$ for each site, $x$, in step 1 (above), one randomly initializes the physical qubit in state $|n_x\>\in \{|0\>,\dots |D-1\>\}$ drawn independently from each site, $x$ from a classical probability distribution: $P(n)$.
Repeatedly sampling from this sto-qMPS ensemble is equivalent to sampling from the mixed state:
\begin{align}
\rho[P] &= \sum_{\vec{n}} P[\vec{n}]  ~|\psi[n_x]\>\<\psi[n_x]|
\nonumber\\
|\psi[\vec{n}]\> &= \sum_{\vec{n}} \ell^T U_A^{n_1',n_1} U_A^{n_2',n_2}\dots|n_1'n_2'\dots \>
\end{align}
where we regard $U_A$ as a four-index tensor with inputs and outputs for physical and bond qubit registers respectively.
Observables, such as $E = \text{tr}(\rho H)$ can be efficiently computed by sampling the sto-qMPS.
The thermal entropy, $S$, of this state is simply the Shannon entropy, $S\(\rho[P]\) = -\sum_{\{n_x\}}P[\vec{n}] \log P[\vec{n}] $, which can be efficiently computed so long as it is possible to efficiently sample from $P$ (or directly compute $S$ analytically).
A graphical representation of a sto-qMPS for general $P$ is shown in Fig.~\ref{fig:MPDO}b.

To formulate a variational approach, we choose $P_{\vec{\pi}}$ to be parameterized by variational parameters $\vec{\pi}$, such that its Shannon entropy,  $S\[P_{\vec{\pi}}\]$, can be efficiently computed $\forall \vec{\pi}$.
This formulation enables efficient computation of the variational free-energy:
\begin{align} 
	F(\vec{\theta},\vec{\pi}) = E(\vec{\theta})-TS(\vec{\pi}).
\end{align}
Here, $E$ is evaluated on a quantum computer, $T$ is specified, and $S$ can (by assumption) be efficiently evaluated classically.
We then seek to minimize $F$ with respect to $\vec{\theta},\vec{\pi}$ to obtain a variational approximation of the true thermal state (see Appendix~\ref{app:opt} for a detailed description of a particular optimization strategy). 
In this way, the sto-qMPS approach variationally approximates a mixed state without doubling the number of physical qubits (as would be required to implement its purification).

Throughout this work, we will restrict our attention to spin-1/2 chains, ($D=2$, $n_i\in \{0,1\}$) with sites indexed by $i=1\dots L$) physical states written in the $S^z$ basis with basis states $s_i = (-1)^{n_i}\in \pm 1$. We will further restrict ourselves to translationally invariant Hamiltonians, $H = \sum_i h_i$ and sto-qMPS ansatzes, so that we can recast the minimization in terms of the free-energy density $f=F/L = \e-Ts$ with energy density: $\e = \<h_i\>$, and entropy density: $s$. 

\subsection{Boltzmann sto-qMPS}
In principle, any probability distribution that can be efficiently sampled classically can be used (e.g. restricted Boltzmann machines, neural networks, et cetera). In this work, we restrict our attention to $P$ that are generated by a classical Boltzmann distribution:
\begin{align}
P[\vec{s}] = \frac{1}{\mathcal{Z}}e^{-W[\vec{s}]},
\label{eq:boltzmann}
\end{align}
where $W$ is a classical (dimensionless) ``energy" function, and the partition function $\mathcal{Z} = \sum_{\vec{n}} P[\vec{s}]$ ensures proper normalization.
To limit the number of additional variational parameters introduced, we further restrict to a simple nearest neighbor interacting form of $W$:
\begin{align}
W = -\sum_i \[Js_is_{i+1}+hs_i\]
\label{eq:clising}
\end{align}
where $J,h$ are (dimensionless) variational parameters. 
A graphical representation of these Boltzmann sto-qMPS is shown in Fig.~\ref{fig:MPDO}c.

Eq.~\ref{eq:clising} includes a special case ($J=0$), the product state ansatz (PSA) introduced in Ref.~\cite{martyn2019product}, in which the initial state of the physical qubit for each site is independently and identically distributed (iid): $P[\vec{s}] = \prod_i \wp(s_i)$ in which case the thermal entropy density is simply $s=S(\wp)$. Formally, the sto-qMPS ensemble with product-spectra corresponds to a two-leg ladder-tensor network.
In the MPS literature, such ladder tensor networks are called local purifications of the MPDO, and have been discussed in the context of open-system MPS calculations~\cite{finsterholzl2020using}.
Incorporating $J\neq 0$ introduces spatial correlations in the sto-qMPS density matrix spectrum, which we will show can surmount certain limitations of the PSA. 

\subsection{Representational power of Boltzmann sto-qMPS}
Generically, the Boltzmann sto-qMPS form (Eq.~\ref{eq:boltzmann}) amounts to replacing the true energy spectrum of the system with that of a classical lattice model. This is clearly not a microscopically faithful representation, as classical Hamiltonians generically lack level repulsion, for example, exhibiting Poissonian level statistics. By contrast, generic quantum Hamiltonians exhibit chaotic (random-matrix-like) level-spacing distributions. Yet, Ref.~\cite{martyn2019product} gave arguments and numerical evidence that physically relevant observables (few-point correlation functions, free-energy density, etc...) cannot be sensitive to the microscopic structure of the Hamiltonian spectrum on energy scales comparable to the level spacing $\delta \sim D^{-L}$, and showed that even a dramatic oversimplification of the spectrum as a product of independent spins, can give a reasonable approximation to local observables. Indeed, this result is very natural from a dynamical perspective: under thermalizing dynamics, starting from a non-thermal state, a region of size $L$ will reach a thermal equilibrium for local observables in time $t\sim L^z$ where $z$ is the dynamical critical exponent (which is finite in thermalizing systems). In contrast, resolving the many-body level spacing requires much longer time: $t_\delta\sim 1/\delta \sim D^L$. Indeed, for real-world systems, $t_\delta$ is typically much longer than the age of the universe so the system cannot have possibly reached an exact Boltzmann distribution on energy scales of order the level spacing! This strongly hints that the details of the many-body spectrum on the scale of $\delta$ are irrelevant to accurately capturing physically-important observables -- a hypothesis we will address empirically through variational optimization for specific spin-chain models.

\section{Variational Thermal State Preparation with sto-qMPS}
\subsection{Model Hamiltonians}
We benchmark the performance of the sto-qMPS variational method by preparing approximate thermal states on a model of quantum magnetism and criticality: the Ising model spin-chain with self-dual (non-integrable) perturbation (SDIM) :
\begin{align}
H_{\text{SDIM}} &= \sum_{i} \[\sigma^x_i \sigma^x_{i+1} + \sigma^z_{i}
    +V  \left(\sigma^z_i \sigma^z_{i+1} + \sigma^x_{i-1} \sigma^x_{i+1} \right) \],
    \label{eq:sdim}
\end{align}
and the Heisenberg chain:
\begin{align}
H_{\text{Heisenberg}} &= \sum_{i}\vec{\sigma}_i\cdot\vec{\sigma}_{i+1}
\end{align}
where $\vec{\sigma}_i$ are Pauli-operators on site $i$. Both of these models are quantum-critical at $T=0$ with power-law correlations and entanglement entropy, $S_E(\ell)$ diverging logarithmically with interval size, $\ell$, as $S_E(\ell) \sim \frac{c}{3}\log \ell$ with $c_\text{SDIM}=1/2$ (for $-2.86<V<250$~\cite{chiu2015strongly}).
Roughly speaking, the amount of entanglement sets the difficulty of representing the ground-state as an MPS, making ground-state of critical systems challenging to capture. 
 At finite $T$, these models become short-range correlated (in accordance with the Mermin-Wagner theorem) with correlations decaying exponentially in distance as $\sim e^{-x/\xi}$ with correlation length $\xi$ that diverges as $T\rightarrow 0$ asymptotically as $\xi\sim T^{-1}$. 
Correspondingly, the $\log$ divergence in the entanglement entropy is cutoff by this thermal correlation length $S_E(\ell\gg T^{-1})\sim \frac{c}{3}\log 1/T$. 
Nevertheless, we continue to focus on models that give diverging entanglement entropy as $T\rightarrow 0$, since i) the finite-T crossover behavior above a quantum critical point or phase is of direct scientific interest, and ii) this regime is governed by an interplay of strong quantum and thermal fluctuations that offers a more challenging test for variational approaches than describing thermal states of gapped models.
 The Heisenberg model is integrable and exactly solvable by Bethe-Ansatz. However, the SDIM is non-integrable for any $V\neq 0$. By comparing the performance for these two models, we will see, for small circuits, that the quality of the qMPDO approximation for thermal states of these models does not depend sensitively on the solvability or central charge (at $T\rightarrow 0$). 
 We will then study the performance of the sto-qMPS ansatz for larger circuit sizes and depths for the SDIM with $V=0$, where the exact solution provides a high-precision numerical benchmark.
%
%

\subsection{Trapped-ion implementation}
\begin{figure}[t]
    \centering
    \includegraphics[width=0.4\textwidth]{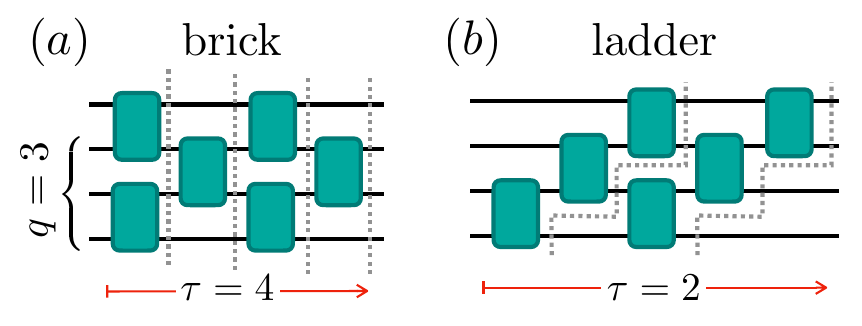}
    \caption{ {\bf Variational circuit architectures -- } Example brick (a) and ladder (b) circuits for number of bond qubits $q=3$, and number of circuit layers $\tau=4,2$ (as indicated in the figure). Circuit layers are bounded by dashed lines. The ladder circuit has depth $\tau(q+1)$ and propagates correlations across all qubits already for a single layer. The brick circuit executes more gates in parallel which can be favorable for experimental implementations in the presence of memory errors.
    }
    \label{fig:circuits_v2}
\end{figure}

As the first test of this method, we explore the variational preparation of thermal states for the spin-chain models described in the last section for various temperatures $T$, using relatively simple circuits with one or two bond qubits. 
The results are both explored through numerical simulations and implemented experimentally using Quantinuum's system model H1 trapped-ion quantum processor, to examine the impact of noise and errors in a real device. 
In the absence of errors, the quality of variational approximation increases with the number of bond qubits, $q$, and circuit depth, $\tau$, used in the qMPS circuit.
For $q = 1$, a single general two-qubit gate ($SU(4)$) provides universal control; for $q = 2$, a (non-universal) depth $\tau=2$ brick-wall circuit with $SU(4)$ gates was used, as represented in Fig. \ref{fig:circuits_v2}.

The complexity of the qMPDO approach turns out to have a non-monotonic dependence on temperature. When $T=0$, a Gibbs's state reduces to a ground state, a pure state that can be described efficiently by single, non-stochastic qMPS. On the other hand, setting $T\rightarrow{}\infty$ returns a maximally mixed state, which also has a trivial representation with $\chi=1$ (i.e. with no bond-qubits). 
We will see, empirically, that the intermediate-temperature regime ($T\sim 1$ in our dimensionless units) between these extreme limits is the most difficult to accurately capture using this approach. 

\paragraph{Implementation Details}
\begin{figure*}
(a)\includegraphics[width=0.45\linewidth]{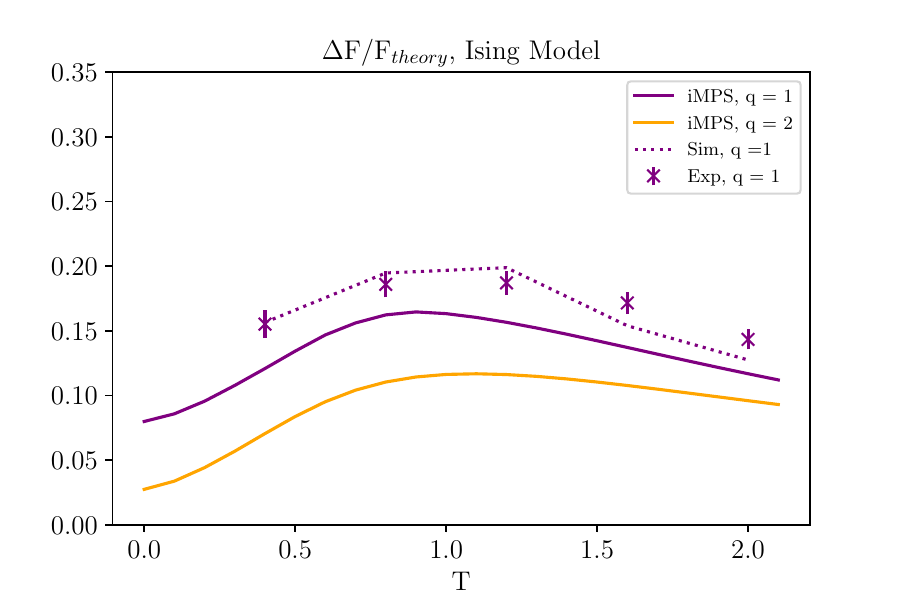}
(b)\includegraphics[width=0.45\linewidth]{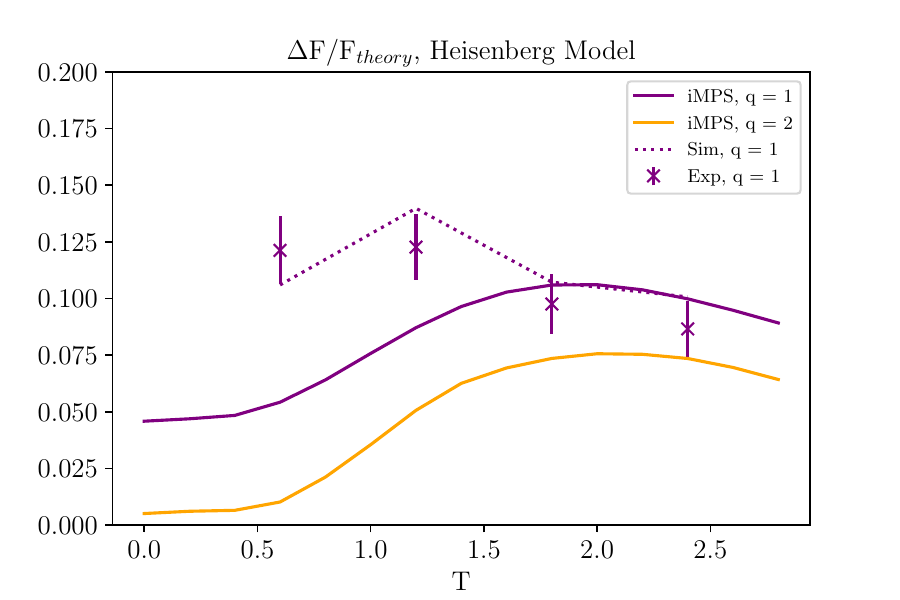}
(c)\includegraphics[width=0.45\linewidth]{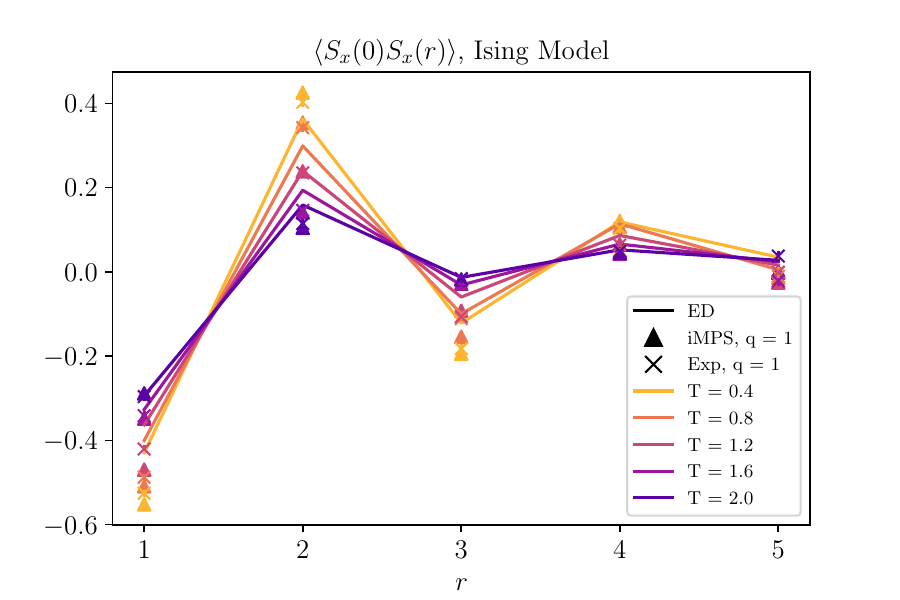}
(d)\includegraphics[width=0.45\linewidth]{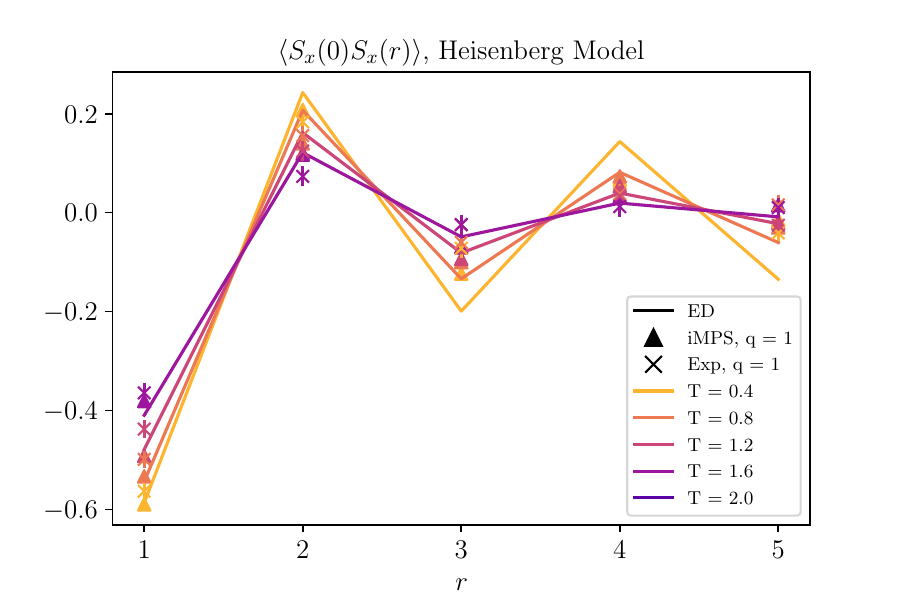}
    
    \caption{ {\bf Small circuit demonstrations of the Ising model -- } {\bf Top:} A temperature scan of the free-energy for SDIM at $V=1$ and Heisenberg model. Quantum hardware results (Exp) are compared with theory, ideal infinite MPS (iMPS), and classical circuit simulations (Sim) using qiskit~\cite{anis2021qiskit} with a featureless depolarizing noise channel with single-qubit (1q) and two-qubit (2q) error rates $\e_{1q}=5\times 10^{-4}$ and $\e_{2q}=8\times 10^{-3}$ respectively. Each data point is averaged over 1,200 and 600 measurement for the SDIM and Heisenberg models respectively. {\bf Bottom:} Correlators were measured for both models at different temperatures. A decrease in correlation length with $T$ can be observed. The sto-qMPS is able to capture correlators for a couple of sites despite the presence of noise.}
\label{fig:HW}
\end{figure*}
In principle, the variational procedure could also be carried out in a hybrid quantum-classical- format, using the quantum processor to compute the estimations of the variational energy (using a large, but finite chain), and a classical optimization loop to tune the variational parameters. To avoid the large sampling overhead associated with this, we instead classically optimize the variational circuit parameters in an infinite system using standard infinite MPS (iMPS) techniques. Specifically, for each set of variational parameters, we compute the corresponding tensors for the ladder tensor network representation of the MPDO (Fig.~\ref{fig:MPDO}), compute the transfer matrix and its dominant eigenvector, and use this as the left environment for measuring the terms in $H$ on three sites (the maximum range of interactions in the SDIM). We then experimentally implement a finite-size version of the the optimized sto-qMPS circuits on Quantinuum's trapped-ion device and compare the results to noisy circuit simulations. For experimental results, the ``bulk'' regime of correlation functions was achieved by ``burning-in'', i.e. iterating the quantum channel for the sto-qMPS for $\ell_\text{burn-in}=5$ sites, which significantly exceeded the spatial correlation length, $\xi\approx 1-3$, of the sto-qMPS for the temperature range explored.


\begin{figure*}
    \includegraphics[width=0.46\textwidth]{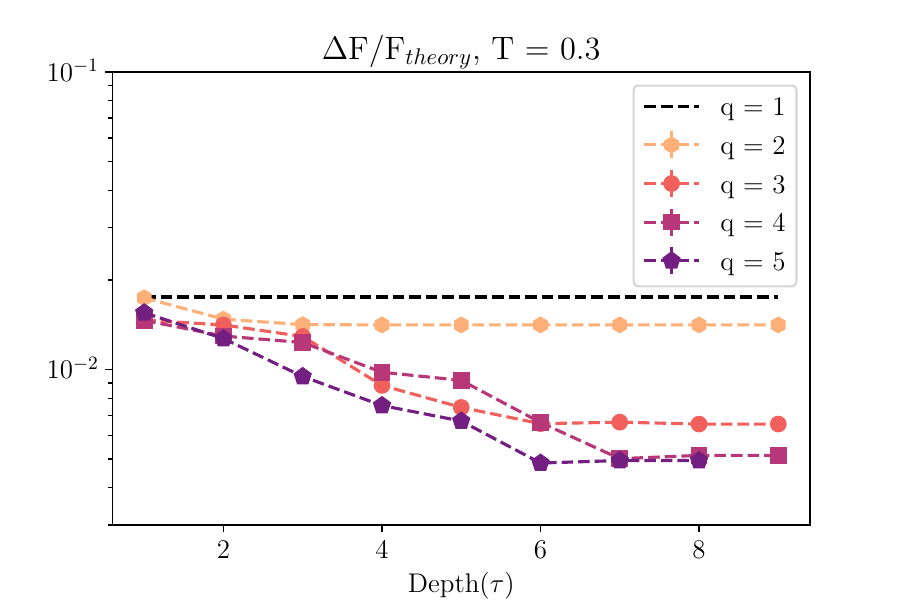}
    \includegraphics[width=0.46\textwidth]{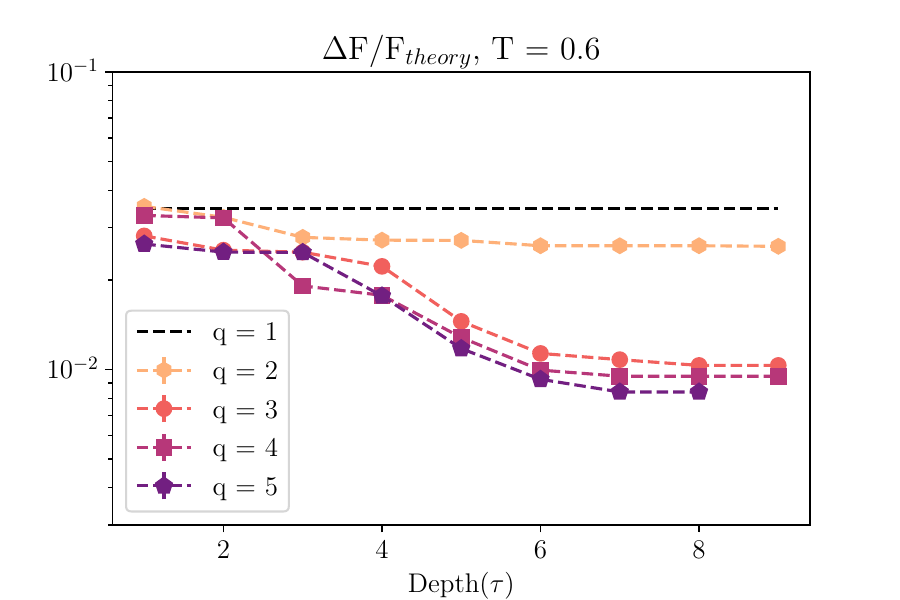}
    \includegraphics[width=0.46\textwidth]{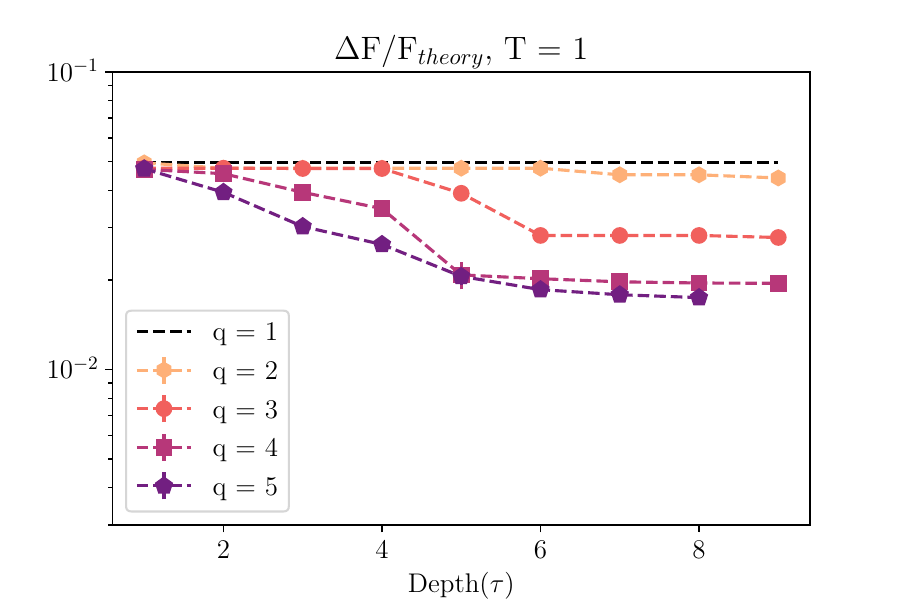}
    \includegraphics[width=0.46\textwidth]{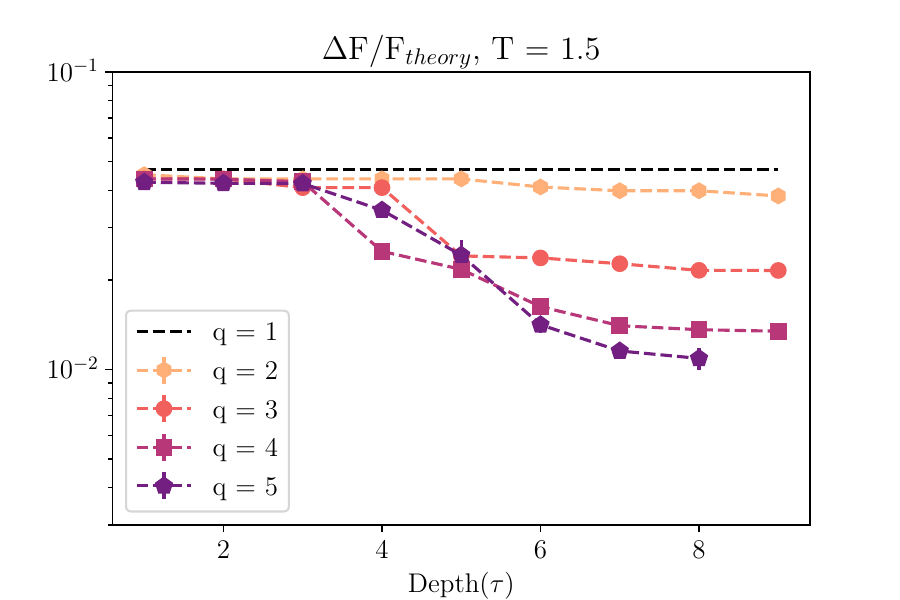}
    \caption{ {\bf Variational power of the thermal sto-qMPS + PSA ansatz -- } The relative error in free-energy, $\Delta F/F$, for the sto-qMPS with a product spectrum ansatz (PSA) and ladder circuit architectures, shown at various temperatures for a length integrable self-dual Ising chain (Eq.~\ref{eq:sdim} with $V=0$), for various number of bond qubits ($q$), and circuit depths ($\tau$) for ladder circuit architectures. $\Delta F/F$ saturates at larger $q$ and $\tau$, which we interpret as an intrinsic limitation of the PSA.
    }    
    \label{fig:quimb}
\end{figure*}

\paragraph{Results for small circuits}
The temperature dependence of the free-energy error and correlation functions for the SDIM is shown in Fig. \ref{fig:HW}, benchmarking against exact diagonalization (ED) simulations on $L=14$ sites with periodic boundary conditions.
In theoretical simulations, the free-energy error decreases with increasing bond-dimension as expected, achieving $\sim 10-15\%$ error with a single bond qubit ($q=1$), and $\sim 5-10\%$ error with a pair of bond qubits ($q=2$).
As a reference, $L=8$ differs from $L=14$ by $\sim 1\%$ so the finite size error in ED is an order of magnitude smaller compared to the sto-qMPS error.

Notice that the error hits a maximum at a moderate temperature, which is at the order of a typical Hamiltonian term's size, and decreases both at low and high temperatures (at very high temperature the error goes to zero quickly as an infinite temperature state is exactly captured by a $q=0$ product state MPDO). We also observe similar non-monotonic behavior for other Hamiltonian parameters ($V$), and in other spin-chain models.

The close agreement between the noisy circuit simulation and experimental results shows that this discrepancy arises mainly from gate errors, whose effects are apparently captured by a simple structureless noise model. 
Thus, achieving higher quantitative accuracy with these methods will require reduction of gate noise, and/or noise mitigation tools, as is often the case for noisy intermediate-scale quantum (NISQ) algorithms. 
 
\section{The variational power of sto-qMPS}
 Previous results have proven that area-law states in 1d are guaranteed to have low bond-dimension MPS representations ~\cite{hastings2007area}.  
Likewise, thermal states in $1d$ are guaranteed to have an area-law for MPDO operator entanglement, and be representable by a low bond-dimension MPDO \cite{kuwahara2021improved}. 
In general, it is not guaranteed that such a low bond-dimension MPDO has an efficient \emph{local} purification; in fact, explicit counterexamples have been constructed~\cite{de2013purifications}. 
Although this obstacle does not apply to ($1d$) thermal states, $\rho\sim e^{-\beta H}$,  where having 
 an efficient (area-law) scaling of operator entanglement for any non-zero temperature~\cite{kuwahara2021improved} 
 enables a local purification: $\rho \sim e^{-\beta H/2}e^{-\beta H/2}$, other problems emerge. For one,
 this local purification is not generally in an appropriate canonical form that can be implemented by a quantum circuit-generated tensor network. Further, the best know approximation algorithm~\cite{kuwahara2021improved} requires a Schmidt rank growing exponentially with the inverse temperature, $\beta$.
In order to address these difficulties with our algorithm, we would also like to empirically understand how the accuracy of a sto-qMPS scales with the quantum memory and circuit resources, to assess the scalability of the sto-qMPS ansatz.
 Here we attempt to address these questions empirically through (simulated) variational optimization of circuit architectures with scalable qubit numbers, $q$ and circuit depths, $\tau$.
 To this end, we focus on the integrable limit of the SDIM (with $V=0$) so that we can compare to the exact answer to obtain a high-precision estimate of errors below achievable finite-size errors with ED simulations.  

\subsection{Product State Ansatz }
The variational power of a given qMPS ansatz is determined by three factors (i) the number of bond qubits, $q$, (ii) the number of circuit layers, $\tau$, and (iii) the internal structure in a local circuit. (i) is related to bond dimension by $\chi=2^q$, (ii) sets on the number of variational parameters per site, and (iii) decides the ``geometry" of our holographic description and thus the rate of entanglement propagation in the virtual bond-space.
Having much freedom at hand, we choose to fix circuit geometry to a ladder circuit (Fig.~\ref{fig:circuits_v2}b) of arbitrary tow-qubit gates and vary $q$ and $\tau$. 
Each incoming leg is connected to either a physical qubit (the top right one) or a bond qubit, $q$ two-qubit gates are then applied on each of the $\tau$ layers. 
Although the ladder geometry is ``denser'' in $\tau$ than other ones such as the multiscale entanglement renormalization ansatz (MERA) or brick-wall, correlations propagate more efficiently in this structure (for a detailed comparison in the qMPS context see \cite{haghshenas2021variational}). 

Since the very-high temperature regime is trivial, and the very low-temperature regime reproduces the ground-state, which is well described by a pure qMPS, we focus on intermediate temperatures $T \in \{0.3,0.5, 1, 1.5\}$, where the small-q simulations above show is the most challenging to describe with a sto-qMPS. Since obtaining an infinite qMPS representation requires exact diagonalization of the transfer matrices, which is computationally intensive for large $q$, we instead simulate a translationally-invariant, finite-size MPS chain and measure correlators in the bulk. 
Specifically, we simulate a semi-infinite chain, using a translation-invariant sto-qMPS ansatz (i.e. using the same variational parameters are used on each site). Energies were measured between sites $48-54$ which reflect bulk behavior~ \footnote{We note that for the semi-infinite geometry averaging the energy over several sites (beyond the range of Hamiltonian terms) was required to avoid pathological configurations where the optimizer exploited the open boundary condition to generate solutions that had low energy density at sites $(i,i+1)$, at the expense of higher overall energy which arose even though the ansatz is translationally invariant.}. Circuits were optimized using the parallel batch method described in Appendix~\ref{app:opt}, with 30 instances per batch, and using the best result for $\tau$ as the initial guess for optimizing $\tau+1$. Optimizations are performed using the tensor network package, quimb~\cite{gray2018quimb}, which is based on a novel tensor network tool, PyTorch\cite{paszke2019pytorch}, and results are shown in Fig.~\ref{fig:quimb}.

Examining the $q$ and $\tau$ dependence, two trends emerge. Overall, the free-energy errors, $\Delta F$, tend to increase with temperature (in accordance with the small-scale $q=1,2$ circuit results above) and decrease with circuit resources $q,\tau$. 
While $\Delta F$ initial decreases with increasing number of circuit layers, $\tau$, this improvement saturates beyond $\tau\approx 4$ for $q=2$ or $\tau\approx 7$ for $q\geq 3$. 
$\Delta F$ also decreases with increasing qubit resources, $q$. 
The saturation value at large $\tau$ initially decreases significantly with $q$ for $q=1,2,3$ but this trend begins to saturate going from $q=3$ to $q=4,5$.

We postulate that this slow-down reflects an intrinsic limitation of the PSA, which may not be overcome even with arbitrary circuit resources $q,\tau\rightarrow \infty$. 
While we cannot completely rule out that the plateau is influenced by local minimum trapping in this high-dimensional space, we view this possibility as unlikely since
%
we are using a sequential optimization routine very similar to that used in Ref.~\cite{haghshenas2021variational} for qMPS ground-state preparation which observed a non-saturating improvement of variational for increasing $q,\tau$ following systematic, non-saturating power-law trends. 
%
This suggests that the saturation error is not set by the complexity of the unitary circuit, but rather by an intrinsic limitation of the PSA that could not be overcome even with arbitrary circuit resources $q,\tau\rightarrow \infty$.
To improve beyond this intrinsic limit, we instead aim at considering a more general class of sto-qMPS that incorporate spatial correlations into the density matrix spectrum.

 \begin{figure}[t!]
    \centering
    \includegraphics[width=0.45\textwidth]{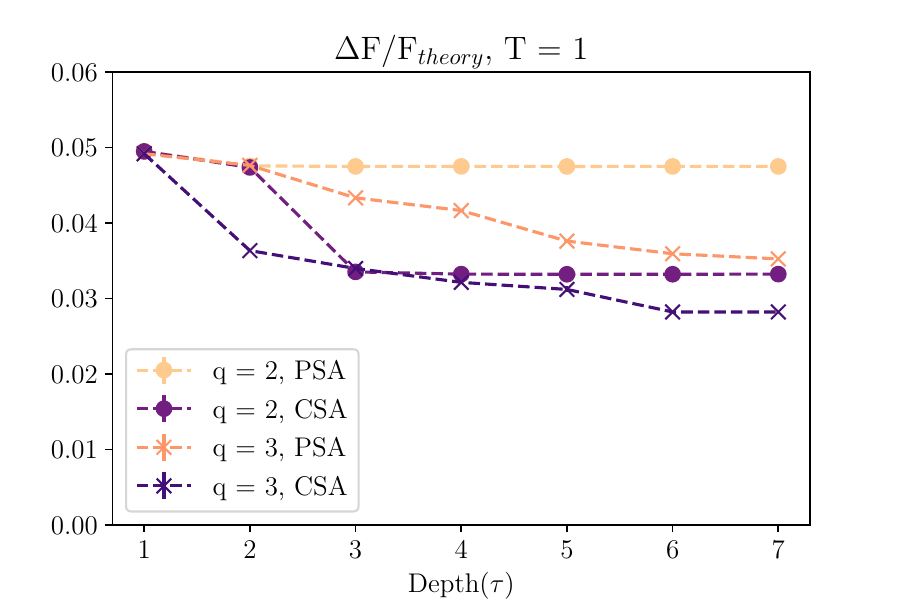}
    \includegraphics[width=0.45\textwidth]{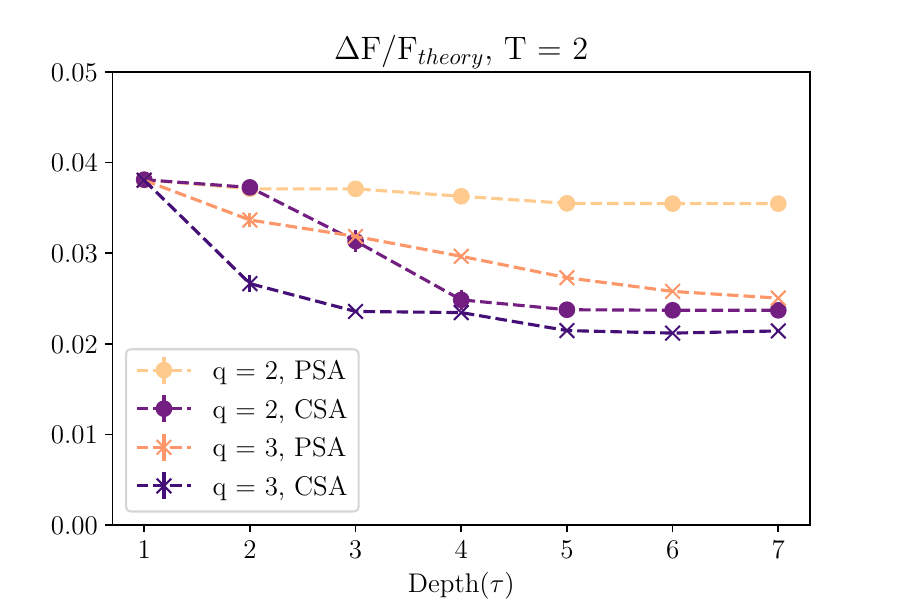}
    \caption{ {\bf Correlated spectrum ansatz} 
    (a) Ladder tensor network corresponding to sto-qMPS with Eq. \ref{eq:clising}.
    We compare the free-energy optimization results at $T = 1$ and $T = 2$ for the critical Ising model, which was previously shown to be one of the most difficult temperatures for PSA in Eq. \ref{eq:clising}.}
        \label{fig:bpsa}
\end{figure}

\subsection{Correlated Spectrum Ansatz (CSA)}
We next explore the use of correlated classical probability distributions, $P$, corresponding to Eq.~\ref{eq:clising} with $J\neq 0$. It is convenient to perform classical calculations by representing the resulting sto-qMPS as a ladder tensor network written in terms of the transfer matrix operator of the Boltzmann weights. We focus on the solvable limit of the SDIM with $V=0$, where the comparison to the exact solution provides high-precision benchmarking. Fig.~\ref{fig:bpsa} shows results for intermediate temperature, $T=1$ and $T=2$ (the most difficult regime for PSA circuits), for the number of bond qubits $q=2,3$~\footnote{Unfortunately, we were not able to use Quimb without modifying the source code for the non-PSA tensor networks, and consequently were restricted to smaller circuits with $q\leq 3$ bond qubits with depths $\leq 7$ due to the slower run time of our non-PyTorch-accelerated custom code}. 
In contrast to the previous section, we choose a translation-invariant ansatz and compute results directly in the infinite system size limit using the technique described in Appendix~\ref{app:bpsa}. 

For all parameters, including spatial correlations into the density matrix spectrum significantly lowers the free-energy error. In fact, the $q=2$ CSA results asymptote with circuit depth, $\tau$, to a similar value to the $q=3$ PSA results, indicating that the single extra variational parameter ($J$) achieves roughly the same improvement as doubling the bond-dimension in the PSA.
We note that many further augmentations of the CSA are possible as targets of future exploration, including increasing the complexity of the classical Hamiltonian, $W$, used to generate $P$ (e.g. including longer-range interactions), adding hidden layers to the weights as in restricted Boltzmann machines, or other more complex classes of probability distributions that could be sampled efficiently (e.g. by Monte Carlo methods).

\section{Discussion}
In this work, we have introduced a qubit-efficient method for preparing thermal states of correlated many-body systems with relatively simple variational circuits. The method is qubit efficient in two senses. First, it uses holographic simulation techniques to represent a thermodynamically large system with a finite number of qubits. Second, by representing a mixed state via stochastically sampling over pure states, it avoids the doubling of qubits required to explicitly implement a purified thermal state. Through classical simulations, we find that relatively simple circuits with only a handful of qubits ($q\approx 3$) can be used to obtain quantitatively accurate (percent- or sub-percent level errors) on the free-energy in the thermodynamic ($L\rightarrow \infty$) limit, which may be sufficient for many practical applications. 

The sto-qMPS works particularly well in relatively low-temperature regimes, relevant to for example studying high-temperature superconductors phenomenology, quantum magnetism, and finite-temperature crossovers above quantum critical points, where the interplay between quantum correlations and thermal correlations is particularly challenging to capture with classical computational techniques; or asymptotically high-temperature regimes relevant to highly non-equilbrium dynamics (even infinite temperature dynamical quantities can still be difficult to compute classically).

The sto-qMPS ansatz can also be particularly helpful in certain quantum dynamics simulations, where a moderately-accurate approximation to an initial thermal state is useful, for example, in simulating the effect of temperature on chemical reaction kinetics or computing the temperature-dependence of conductivity in a correlated electron system. In these applications, the qubit-efficient sto-qMPS ansatz can serve as an initial state for qubit-efficient holographic quantum dynamics techniques~\cite{chertkov2021holographic} to simulate near-thermal equilibrium dynamics. In this context, generic dynamics will be thermalizing and quickly erase any small errors in the initial state (yet it is still important to be able to produce a state with a known, and tunable initial temperature).

These preliminary explorations of sto-qMPS leave several open directions for future study.
 For example, it may be advantageous  explore more complex classical probability distributions such as Boltzmann weights with longer-range interactions or models with hidden layers such as restricted Boltzmann machines or neural network parameterizations.
Another important challenge is to explore the scalability of these techniques to higher dimensional systems, for example, generalizing the MPS structure to a stochastic (isometric~\cite{zaletel2020isometric}) tensor network state (sto-qTNS) whose tensors are isometries generated by quantum circuits acting on bond-qubits and stochastically generated initial states of physical qubits.
Here, classical computation will become increasingly demanding but may be tractable for modest system sizes to explore the scalability of these ideas to $2d$ systems.

\vspace{4pt}\noindent{\it Acknowledgements -- }
We thank Tomotaka Kuwahara for helpful discussion, and Garnet Chan, Michael Foss-Feig, David Hayes, Shyam Shankar, and Mike Zaletel for insightful conversations and previous related collaborations. This work was supported by NSF Award DMR-2038032 and the Alfred P. Sloan Foundation through a Sloan Research Fellowship (ACP). This research was undertaken thanks, in part, to funding from the Max Planck-UBC-UTokyo Center for Quantum Materials and the Canada First Research Excellence Fund, Quantum Materials and Future Technologies Program.

\bibliography{qmpdobib.bib}

\appendix

\section{Implementation details}
Hardware implementations were performed on Quantinuum's system model H1 trapped-ion quantum processor, which is based on a quantum charge-coupled device (QCCD) architecture and uses hyperfine clock states of $\rm ^+Yb^{171}$ qubit ions, and laser-based gates~\cite{Pino2021}. Hardware-native gates include arbitrary single-qubit gates, and entangling two-qubit ``ZZ-gate" operation $U_{2q} = e^{i\pi/2 \sigma^z\otimes \sigma^z}$. 


Due to the low clock rate and large sampling cost, we did not perform the variational optimization on the quantum hardware, but rather pre-optimized circuit parameters classically and implemented the optimized circuits. The results test the impact of realistic hardware noise, which appears to match simulations with the simple featureless depolarizing noise model.

\begin{figure}
\includegraphics[width=0.4\textwidth]{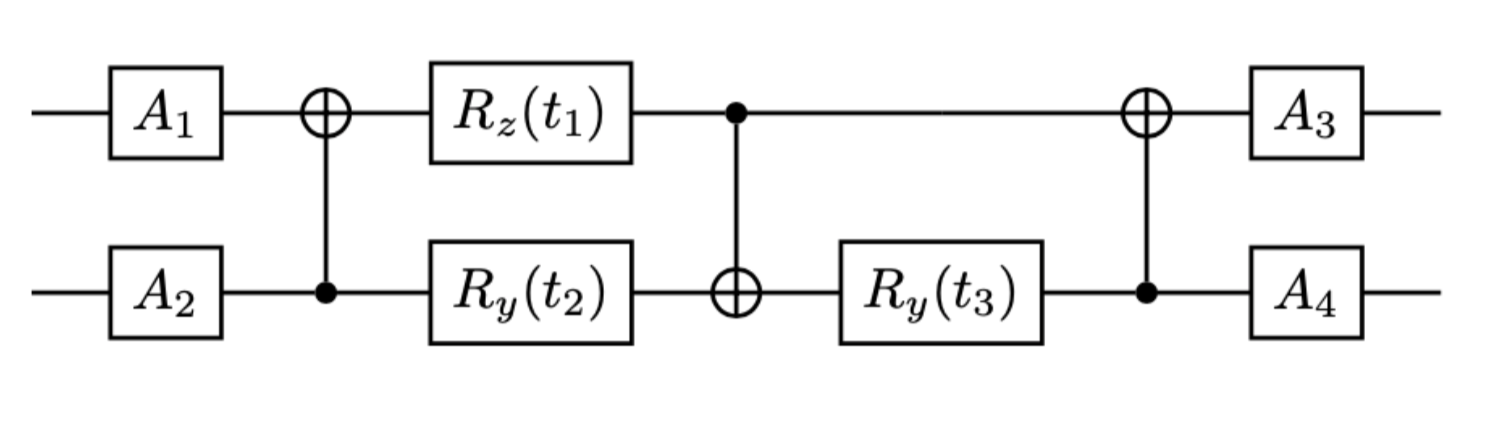}
\label{appfig:su4}
\caption{{\bf Circuit decomposition for $SU(4)$ gates} $R_{\alpha=x,y,z}(\theta) = e^{-i\frac{\theta}{2}\sigma^\alpha}$, and the CNOTs are compiled from the native 2q gate by dressing the native ZZ-gate with 1q gates. $A$ denote arbitrary single qubit gates which can be compiled from alternating sequence of $R_x$, $R_z$, and $R_x$ gates. }
\end{figure}

Our quantum circuit generated tensors are based on general two-qubit gates realizing arbitrary $SU(4)$ unitary operations. We used the circuit decomposition shown in Fig.~\ref{appfig:su4}. 
Another implementation detail is that to reduce the number of distinct circuits that must be compiled, we used a quantum random number generator to randomly initialize the physical qubits. Specifically, using an ancilla qubit, we prepared the entangled state $\sqrt{p}|00\>+\sqrt{1-p}|11\>$, and measured the second qubit (ancilla) qubit, to achieve a mixed state $p|0\>\<0|+(1-p)|1\>\<1|$ for the physical qubit, where $p\in [0,1]$.

We did not employ any error mitigation strategies, though standard methods of error mitigation could readily be incorporated into this technique to improve the performance of noisy hardware.

%

\section{Batch optimization heuristics \label{app:opt}}
Globally optimizing a large tensor network is computationally challenging, and it may generally suffer from barren plateau problems where the gradient of the objective function nearly vanishes \cite{cerezo2021cost}; on the other hand, a local optimization, such as the Limited-Memory Broyden–Fletcher–Goldfarb–Shanno (L- BFGS-B) algorithm, gives results heavily dependent on the initial conditions, and may become trapped in local minima with high free-energy error. To overcome this dilemma, we adapted a batch-sequential optimization strategy in our numerical studies. 
Specifically, we begin with a batch of $N_\text{batch}$ different single-layer circuits, each with random initial parameters.
For iMPS we used the above parameterized circuit representation of the general $SU(4)$, but for the GPU-accelerated finite-MPS calculations (large circuits), it was more convenient to directly work with the matrix entries in the computational basis. Then to generate randomness, we added a random number to each parameter i.i.d. $\sim [0,x]$ where $x$ characterizes the strength of randomness. For iMPS calculations, the randomness was added directly to the circuit parameterize. For finite MPS calculations, the randomness was added directly to each entry of the unitary matrix followed by a QR factorization to restore unitarity.

Each circuit in the batch is optimized (in parallel) with a local optimizer, and the best outcome of the batch is selected.
This best-of-batch example is used to generate another batch of $N_\text{batch}$ circuits with $\tau=2$, by adding another gate layer of identity and then some small randomness to all gate parameters. Our method not only preserves the good qualities of the optimized first layer but also gives a chance to get kicked out of a local variational free-energy minimum.
\begin{figure}
\centering
\includegraphics[width=0.50\textwidth]{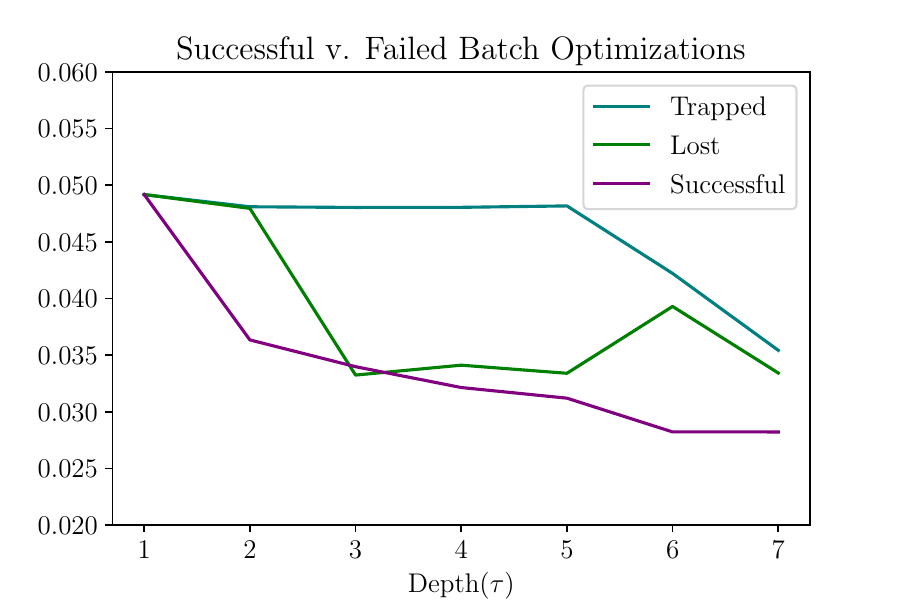}
\label{appfig:batch}
\caption{{\bf Randomness in the batch optimization method} Two unsuccessful $q = 3$ optimizations compared with the successful CSA $T=1$ data shown in Fig. \ref{fig:bpsa}. If too much randomness is added to the iterative batch optimization, the optimizer of the new iteration will lose its memory from the previous best run (``lost"). On the other hand, too little randomness tends to make the optimizer trapped (``trapped"). The randomness parameter was chosen to produce behavior labeled here as ``successful", where the optimization decreases monotonically with circuit depth without trapping. }
\end{figure}
\begin{table}[t]
  \begin{center}
    
    \label{tab:table1}
    \begin{tabular}{l |c*{1}{c}|l|c*{5}{c}} 
      \textbf{iMPS} & \textbf{$q = 2$} & \textbf{$q = 3$}& \textbf{finite MPS}& \textbf{$q = 2$} & \textbf{$q= 3$} &\textbf{$q= 4$} &\textbf{$q= 5$}  \\ 
      \hline
     $\tau \leq 4$ & 0.5 & 0.3 &$\tau \leq 4$ &0.2 &0.12 &0.1 &0.07 \\
     $\tau > 4$ & 0.4 & 0.2 &$\tau >  4$ &0.2 &0.12 &0.08 &0.05 \\ 
    \end{tabular}
    \caption{\textbf{Batch randomization hyperparameter} To optimize circuits with $\tau+1$ layers, we took a batch of samples with the optimal parameters of the $\tau$-layer circuits. For iMPS, we perturbed the circuit parameters by a value i.i.d.$\sim [0,x]$ for each circuit parameter, whereas in the finite chain case, we added a random amount i.i.d.$\sim [0,x]$ to each element of the tensor network and performed QR decomposition to restore unitarity. Values of $x$ are listed in the table for various $q,\tau$.}
  \end{center}
\end{table}
Each element of this batch of $\tau=2$ circuits is then optimized in parallel (also allowing the thermal parameter, $p$ to vary), and the one with the lowest free-energy is used to generate a batch of $\tau=3$ circuits, and so on.
The procedure is repeated with parallel batches until the target depth circuit is reached. Depending on the number of parameters in the circuit, one needs to adjust the amount of randomness accordingly: more parameters require less randomness. Too much randomness can result in failure of the optimization, as shown in Fig. \ref{appfig:batch}.

\section{Classical tensor network simulations of sto-qMPS with correlated spectra \label{app:bpsa}}
Any local classical Hamiltonian $W$ generating the probability distribution $P$ on the physical qubit initial states (see Eq.~\ref{eq:boltzmann}) can be written in terms of a transfer matrix. For numerical simulations, it is convenient to implement that transfer-matrix as a (diagonal) MPO, as shown in Fig.~\ref{fig:MPDO}.
We note that, for small circuits, standard methods can be used to compute the free-energy directly in the infinite system size limit (i-sto-qMPS), by properly normalizing the transfer matrix to have the largest eigenvalue of $1$, and projecting the left boundary vector of the MPO for $P$ onto the corresponding eigenvector.

For the nearest-neighbor classical Ising model, the transfer matrix eigenvectors, and corresponding entropy can be readily computed analytically.
For convenience, we generalize Eq.~\ref{eq:boltzmann} to include an inverse temperature, $\beta$: $P_\beta = \frac{1}{\mathcal{Z}_\beta}e^{-\beta W}$, which we will take to unity at the end of the calculation. The partition function $Z_\beta$ can be written in terms of the transfer matrix $Z_{\beta}= \<\vec{1}|\(T_{\beta}\)^{L}|\vec{1}\>$ where $|\vec{1}\>$ is the vector with all unit entries, and: 
\begin{align}
T_{\beta} = \begin{pmatrix}
e^{\beta (J+h)} & e^{-\beta J}\\
 e^{-\beta J}& e^{\beta (J-h)}
\end{pmatrix}
\end{align}

Diagonalizing $T_\beta$ gives eigenvalues:
 \begin{align}
 \lambda^\pm_{\beta} =e^{\beta J}\text{cosh}(\beta h)\pm\sqrt{e^{2\beta J} \text{sinh}^2(\beta h) + e^{-2\beta J}}
 \end{align}
 and $Z_\beta = |a_+|^2\lambda_+^L+|a_-|^2 \lambda_-^L$, where $a_\pm$ are the coefficients of $\vec{1}$ in the eigenbasis of $T_\beta$. For large $L$, only the dominant eigenvalue $\lambda_+$ contributes.
 Entropy density is then given explicitly as: 
 \begin{align}
 s = \frac{1}{L}\lim_{\beta\rightarrow 1} (-\frac{\d}{\d\beta}+1)\text{log}Z_{\beta}
 \end{align}

The method can be readily generalized to more complicated $W$. For example, if $W$ has $k$-nearest neighbor interaction terms, where the transfer matrix is a $2^k\times 2^k$ matrix, and the rest of the calculation follows similarly.
 
\end{document}